\documentclass[amssymb,prb,aps,twocolumn,amsmath,showpacs]{revtex4}

\usepackage[dvips]{graphicx}
\usepackage{amstext}

\begin{document}

\title{Observation of standard spin-switch effects in F/S/F trilayers with a strong ferromagnet}
\author{Ion C. Moraru, W. P. Pratt, Jr., Norman O. Birge}
\email{birge@pa.msu.edu}
\affiliation{Department of Physics and
Astronomy, Michigan State University, East Lansing, Michigan
48824-2320, USA}
\date{\today}

\begin{abstract}
We have measured the superconducting transition temperature $T_c$
of F/S/F trilayers using Permalloy (Py=Ni$_{84}$Fe$_{16}$) as a
strongly polarized ferromagnetic material.  For a parallel (P) or
anti-parallel (AP) alignment of the magnetization directions of
the outer ferromagnets, we observe a $T_c$ difference as large as
20 mK, with a stronger suppression of superconductivity in the P
state than in the AP state. This behavior is opposite to the
recent observations of Rusanov \textit{et al.}, Phys. Rev. B
\textbf{73}, 060505 (2006) in Py/Nb/Py trilayers, but is
consistent with earlier results on trilayers with Ni or CuNi alloy
as the ferromagnetic material.

\end{abstract}

 \maketitle

The presence of a ferromagnetic (F) material in contact with a
conventional superconductor (S) results in a strong mutual
influence.\cite{Izyumov2002}  The superconducting correlations
penetrate into the ferromagnet and oscillate in sign over a very
short distance, due to the large energy difference between the
majority and minority spin bands in the ferromagnet. Bilayers,
trilayers, and multilayers of S and F materials exhibit a wide
variety of novel phenomena, including oscillations of the
superconducting critical
temperature\cite{ChienReich,Mercaldo1996,Muhge1996} and density of
states\cite{KontosPRL2001}, and Josephson junctions with a
$\pi$-shifted ground state.\cite{RyazanovPRL2001}

In this paper we focus on the so-called ``superconducting spin
switch" first discussed in 1966 by deGennes\cite{deGennes1966} and
rediscovered in 1999 by Tagirov\cite{TagirovPRL1999} and by
Buzdin, Vedyayev and Ryzhanova.\cite{BuzdinEPL1999} Those authors
predicted that the critical temperature, $T_c$, of a F/S/F
trilayer should depend on the relative magnetization direction of
the two F layers, with the smallest $T_c$ occurring in the
parallel (P) state and the largest $T_c$ in the antiparallel (AP)
state.  Those predictions were verified long ago by Deutscher and
Meunier,\cite{DeutscherMeunier1969} and more recently by Gu
\textit{et al.},\cite{GuPRL2002} Potenza and
Marrows,\cite{PotenzaPRB2005} and Moraru \textit{et
al.}\cite{MoraruPRL2006} in a variety of F/S/F systems. It came as
a surprise, therefore, when Rusanov \textit{et al.}
\cite{RusanovPRB2006} recently reported observation of the
\textit{inverse} spin switch effect in a series of Py/Nb/Py
trilayer samples.  Although the difference in $T_c$ between the P
and AP magnetization configurations was small in that work, the
data showed clearly that the resistance in the transition region
was higher for the AP configuration than for the P one.  In fact,
similar behavior had previously been observed by Pe\~{n}a
\textit{et al.}\cite{PenaPRL2005} in F/S/F trilayers made from
superconducting YBa$_2$Cu$_3$O$_4$ and ferromagnetic
La$_{0.7}$Ca$_{0.3}$MnO$_3$, with a spin polarization expected to
be close to 100\%.  Those authors interpreted their observations
as arising from enhanced reflection of spin-polarized
quasiparticles at the F/S interfaces in the AP state leading to a
stronger suppression of superconductivity,\cite{Takahashi1999} and
Rusanov \textit{et al.}\cite{RusanovPRB2006} claimed that the
inverse spin-switch behavior is generic for F/S/F trilayers with
strong ferromagnets. We believe that the mechanism based on
reflection of quasiparticles at the S/F
interface\cite{Takahashi1999} can explain changes in resistance
under nonequilibrium conditions, but cannot explain differences in
the equilibrium $T_c$ between the P and AP states. Given our
earlier work showing standard spin-switch behavior in Ni/Nb/Ni
trilayers,\cite{MoraruPRL2006} we were motivated to carry out
independent measurements of $T_c$ in Py/Nb/Py trilayers.

A series of Py(8)/Nb($d_{s}$)/Py(8)/Fe$_{50}$Mn$_{50}$(8)/Nb(2)
multilayers (all thicknesses are in nm) was fabricated, with
thicknesses for the superconducting layer, $d_s$, varying between
20 and 150 nm. The samples were grown directly onto Si substrates
by magnetically-enhanced triode dc sputtering in a high vacuum
chamber with a base pressure in the low $10^{-8}$ Torr and an Ar
pressure of $2.0\cdot10^{-3}$ Torr. The thickness of the Py layers
were chosen to be much longer than the dirty-limit coherence
length, $\sqrt{\hbar D_F^{\uparrow}/E_{ex}}$, which we estimate to
be 1.2 nm for the majority spin band using $E_{ex}$ = 0.135
eV,\cite{PetrovykhAPL1998} $D_F^{\uparrow} = v_F^{\uparrow}
l_F^{\uparrow}/3$, $v_F^{\uparrow} = 0.22\cdot10^{6}$ m/s and
$l_F^{\uparrow}=4$ nm.\cite{AltmannPRL2001} The FeMn layer fixes
the direction of the top Py layer by exchange bias
\cite{NoguesJMMM1999} after undergoing a brief annealing and
in-field cooling process. The Nb capping layer protects the FeMn
from oxidation and is not superconducting.

Samples were patterned for four-terminal current-in-plane
resistance measurements by mechanical masking during sputtering.
The lateral dimensions of the samples were 4.3 mm x 1.6 mm.  The
critical temperatures of all samples were determined by ac
resistance measurements with current of 10 $\mu$A, corresponding
to a current density less than $3 \times 10^5$ A/m$^2$, low enough
to be in the linear response regime. $T_c$ was defined to be the
temperature at which the resistance dropped to half its normal
state value.
\begin{figure}[ptbh]
\begin{center}
\includegraphics[width=3.2in]{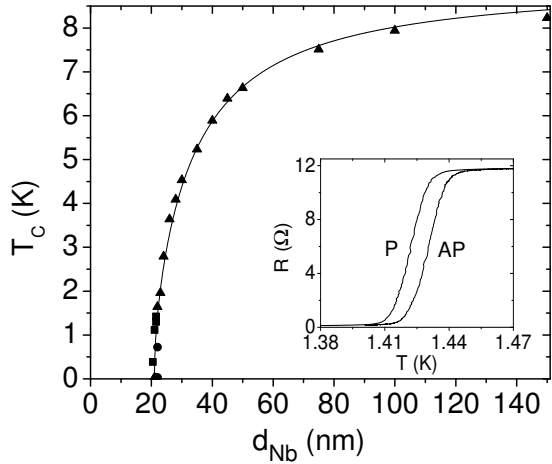}
\end{center}
\caption{Critical temperature vs. Nb thickness for a series of
Py(8)/Nb($d_s$)/Py(8)/Fe$_{50}$Mn$_{50}$(8)/Nb(2) samples (all
thicknesses are in nm). The various symbols represent different
sputtering runs. The solid line represents the theoretical fit as
explained in the text. Inset: R vs. T for a $d_s$=21.5 nm sample
illustrating the difference between $T_c$ for the P and AP
states.} \label{Fig1}
\end{figure}

\begin{figure}[ptbh]
\begin{center}
\includegraphics[width=3.2in]{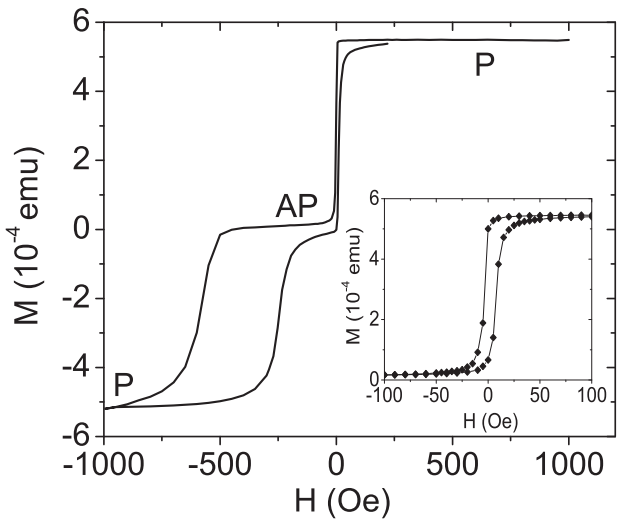}
\end{center}
\caption{Magnetization vs. applied field for a $d_s = 23$ nm sample
measured at $T = 4.2$ K. At $H\approx\pm$10 Oe the free bottom Py
layer switches while the pinned top Py layer switches at around -500
Oe. Inset: minor loop measured at $T = 4.2$ K showing good switching
of the free Py layer.} \label{Fig2}
\end{figure}
The results for the $T_c$ measurements are summarized in Fig. 1.
The $T_c$ of the trilayer shows a strong dependence on the
superconductor thickness close to a critical thickness,
$d_s^{cr}$, where the sensitivity to ferromagnetism is enhanced.
No superconductivity is observed above 36 mK for $d_s <
d_s^{cr}\approx 20.5$ nm.

We have verified the magnetic configuration of our structures on
simultaneously sputtered samples of larger lateral size using a
SQUID magnetometer.  Fig. 2 shows a plot of magnetization vs.
applied field, $H$, for a sample with $d_s$ = 23 nm taken at 4.2
K, illustrating the typical spin-valve behavior of the trilayer.
The narrow hysteresis loop near $H$ = 0 shows the switching of the
free Py layer with a coercive field $H_c = 5-10$ Oe, while the
wider loop shows switching of the pinned layer, shifted to nonzero
$H$ due to the exchange bias with the FeMn.  The inset to Fig. 2
shows a minor hysteresis loop illustrating that applied fields of
$\pm$100 Oe switch the trilayer fully between the P and AP
configurations. The nearly zero net magnetization observed at -100
Oe suggests very good AP alignment, while the nearly saturated
magnetization observed at +100 Oe indicates good P alignment.
Similarly, well-defined alignment of the P and AP states can be
achieved at temperatures in and below the superconducting
transition.

Measurements of $T_c^P$ and $T_c^{AP}$ were performed by
alternating the applied field between +100 and -100 Oe, while the
temperature was slowly decreased through the transition region.
The largest shift in critical temperature, $\Delta T_c \equiv
T_c^{AP} - T_c^P$, should occur in samples with the Nb thickness
close to $d_s^{cr}$. The inset to Fig. 1 shows a plot of $R$ vs.
$T$ for a sample with a nominal thickness $d_s$ = 21.5 nm,
measured in a dilution refrigerator, with a $T_c=1.42$ K.  Two
distinct transitions are observed for P and AP alignment, with a
separation in temperature $\Delta T_c \approx 9$ mK for this case.
Samples with $d_s \approx 22$ nm have $T_{c}$'s between 2 and 3 K
and exhibit values for $\Delta T_c$ of only a few mK, similar to
results obtained previously in other F/S/F
systems.\cite{GuPRL2002,PotenzaPRB2005,MoraruPRL2006} There is no
observable difference between the P and AP state for samples with
$d_s>26$ nm.

Figure 3 shows a plot of $\Delta T_c$ vs. $T_c$ for nine samples.
The largest observed $\Delta T_c$ for our Py/Nb/Py trilayers is
about 20 mK for a sample with $d_s=20.5$ nm and $T_c=0.385$ K. The
data are somewhat scattered for the thinnest Nb layers due to the
increased sensitivity of $T_c$ to small variations of thickness
and growth conditions. Nonetheless, our samples always show that
$T_c^P < T_c^{AP}$, a result that is opposite to what was observed
by Rusanov \textit{et al.}\cite{RusanovPRB2006} in similar
Py/Nb/Py trilayer systems.

\begin{figure}[ptbh]
\begin{center}
\includegraphics[width=3.2in]{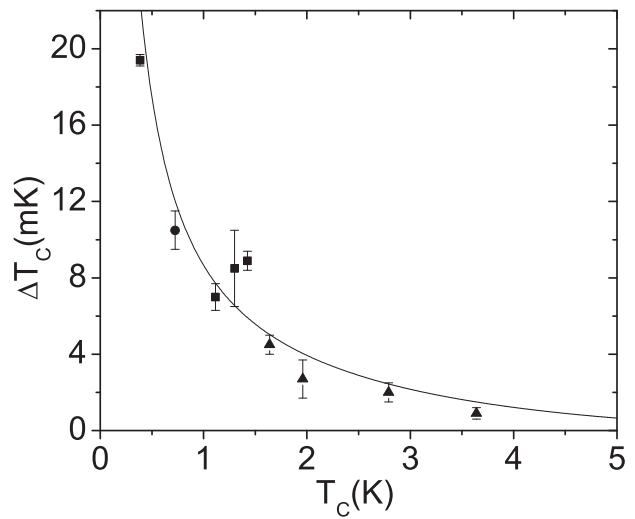}
\end{center}
\caption{$\Delta T_c$ vs. critical temperature for a series of
Py(8)/Nb($d_s$)/Py(8)/Fe$_{50}$Mn$_{50}$(8)/Nb(2) illustrating the
difference in $T_c$ between the P and AP state.  The fit to the
data is obtained using the theory of Fominov \textit{et
al.}\cite{FominovJETP2003} as outlined in the text.} \label{Fig3}
\end{figure}

The critical temperature of F/S/F trilayers in the P and AP states
has been calculated theoretically by several groups.
\cite{TagirovPRL1999,BuzdinEPL1999,FominovJETP2003,BaladiePRB2003,YouPRB2004}
The usual approach involves solving the Usadel equations in the
dirty limit, which for the superconductor implies $l_S < \xi_{BCS}
= \hbar v_S \gamma/\pi^2 k_{B}T_{c\,0}^b$, and for the ferromagnet
$l_F < \hbar v_F/E_{ex}$, where $l_S$ and $l_F$ are the electron
mean free paths in S and F.  Here, $T_{c\,0}^b$ is the transition
temperature of the bulk superconductor, $v_S$ and $v_F$ are the
Fermi velocities in the S and F materials, and $\gamma=1.7811$.
These simplified theories do not consider different electronic
properties for the majority and minority spin bands of the F
material.

We compare our data with the theory of Fominov \textit{et
al.}\cite{FominovJETP2003}  The following equations, which
describe the critical temperatures $T_c^P$ and $T_c^{AP}$ for the
P and AP cases, are obtained in the limit of a thin S layer with a
constant superconducting gap $\Delta$, and a strong ferromagnet
with $E_{ex} \gg \Delta$:
\begin{equation} \label{TcEqnFominovP}
\ln\,\frac{T_{c\,0}}{T_c^P} - Re\,\Psi\left(\frac{1}{2} +
\frac{V_h}{2}\frac{\xi_S}{d_s}\frac{T_{c\,0}}{T_c^P}\right)+\Psi\left(
\frac{1}{2}\right)=0
\end{equation}
\begin{equation}\label{TcEqnFominovAP}
\ln\,\frac{T_{c\,0}}{T_c^{AP}} - \Psi\left(\frac{1}{2} +
\frac{W}{2}\frac{\xi_S}{d_s}\frac{T_{c\,0}}{T_c^{AP}}\right)+\Psi\left(
\frac{1}{2}\right)=0,
\end{equation}
where $\xi_S=\sqrt{\hbar D_S / 2\pi k_B T_{c\,0}}$ and $T_{c\,0}$
is the critical temperature for an isolated superconducting layer
of thickness $d_s$.  Fominov \textit{et al.} make the important
point that the existence of a significant dependence of $T_c$ as a
function of the relative magnetization angle for $d_s>\xi_S$ is
due to the fact that the critical temperature of the trilayer is
suppressed as compared to that of the isolated Nb layer, i.e. $T_c
\ll T_{c\,0}$. Consequently, the condition for which this theory
is valid, $d_s \ll \xi = \sqrt{\hbar D_S/(2\pi k_B T_c)}$, is
considerably weaker than the condition $d_s\ll\xi_S$, because $\xi
\gg \xi_S$.  In the limit of thick ferromagnets, the \textit{tanh}
functions in ref. \cite{FominovJETP2003} are set to 1, and the
functions $V_h$ and $W$ in Eqns. \ref{TcEqnFominovP} and
\ref{TcEqnFominovAP} become
\begin{equation} \label{VhEqnRed}
V_h=\frac{\rho_S\xi_S}{(1-i)\rho_F/2k_h+R_BA} \qquad W=Re\{V_h\}
\end{equation}
where $k_h=\sqrt{E_{ex}/\hbar D_F}$ is the inverse coherence
length in F and $R_BA$ is the boundary resistance times area of
the S/F interface, a parameter whose value reflects both the
quality of the interface and the Fermi surface mismatch between
the S and F materials. Eqns. (1-3) produced the fits to the
Py/Nb/Py data shown in Fig. 1 and 3.

Estimates for the parameters appearing in the theory were obtained
by performing additional measurements on bulk and thin film
samples.  Since the F layer is treated in the thick limit and its
thickness remains fixed for all our samples, we have used a bulk
value for the resistivity of Py, namely $\rho_F=123$ n$\Omega$ m.
\cite{PrattAPL1996}  By contrast, the thickness of the S layer in
our trilayers changes, and thus we have measured $\rho_S$ as a
function of the thickness on bare Nb thin films.  In addition, the
variation of $T_{c\,0}$ with thickness was also measured on the
same Nb films.  The explicit dependencies of $\rho_S$ and
$T_{c\,0}$ were taken into account in Eqns. \ref{TcEqnFominovP}
and \ref{TcEqnFominovAP}. \cite{ThicknessDependence}    The
coherence length was obtained by performing perpendicular field
measurements on the bare Nb films, giving $\xi_S\approx 6$ nm in
the thickness range of our data. Taking the limit of $T_c
\rightarrow 0$ in Eq. \ref{TcEqnFominovP}, for the behavior as
$d_s$ approaches the critical thickness $d_{s}^{cr}$, results in
the relation $d_{s}^{cr}/\xi_S=2e^C| V_h|$ where $C=0.577$ is the
Euler constant.  Using this constraint in Eq. \ref{VhEqnRed} one
can obtain an estimate for the boundary resistance:
\begin{equation}\label{FominovConstr}
R_BA\approx2e^C\rho_S(d_s^{cr})\frac{\xi_S^2}{d_s^{cr}}
\end{equation}
where the value of $\rho_S$ is taken at the critical thickness.
After constraining $R_BA$ as shown above and using the measured
values for the resistivities and $\xi_S$, $k_h$ is the only
remaining fit parameter.

Using Eq. \ref{FominovConstr} with $d_s^{cr}=20.5$ nm and
$\xi_S=6$ nm gives the estimate $R_BA=1.5$ f$\Omega$ m$^2$, which
when utilized in Eq. \ref{TcEqnFominovP} yields a fit that follows
the $T_c$ vs $d_s$ data very well, as shown in Fig. 1. The fit to
the $T_c$ vs $d_s$ data is somewhat insensitive to the value of
$k_h$, however. That parameter is tightly constrained by fitting
to the $\Delta T_c$ vs $T_c$ data.  The results of that fit are
illustrated in Fig. 3, showing good agreement to the data with
$k_h=1.0$ nm$^{-1}$. In comparison, independent estimates of $k_h$
using the values of $v_F$ and $l_F$ discussed earlier are $0.8$
and $2.0$ nm$^{-1}$ for the majority and minority spin bands,
respectively.\cite{AltmannPRL2001}

\begin{figure}[ptbh]
\begin{center}
\includegraphics[width=3.2in]{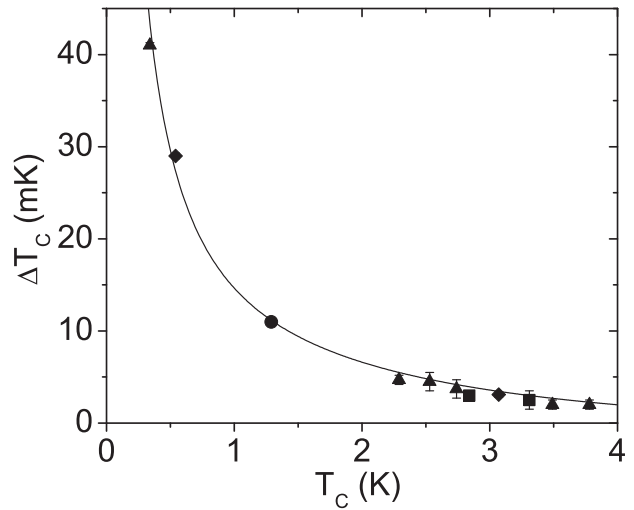}
\end{center}
\caption{$\Delta T_c$ vs. critical temperature for a series of
Ni(7)/Nb($d_s$)/Ni(7)/Fe$_{50}$Mn$_{50}$(8)/Nb(2).\cite{MoraruPRL2006}
The fit to the data is obtained using the theory of Fominov
\textit{et al.}\cite{FominovJETP2003}} \label{Fig4}
\end{figure}

The excellent fit shown in Fig. 3 motivated us to apply the theory
of Fominov \textit{et al.} to our previously-reported data on
Ni/Nb/Ni trilayers.\cite{MoraruPRL2006} The results of the fit to
the $\Delta T_c$ vs $T_c$ data from the Ni/Nb/Ni trilayer are
illustrated in Fig. 4, and also show excellent agreement, even
though it is not obvious \textit{a priori} that the Ni layers in
those samples are in the dirty limit. The values $\rho_F=33$
n$\Omega$m, $\xi_s=6$ nm and $d_s^{cr}=16.5$ nm were used in the
fit, which gave $R_BA=2.3$ f$\Omega$m$^2$ for the Ni/Nb interface
and $k_h=0.5$ nm$^{-1}$. We have made independent measurements of
the Nb/Ni interface resistance using
current-perpendicular-to-plane resistance measurements of Nb/Ni
multilayers, and find the value $R_BA=2.35\pm0.25$ f$\Omega$
m$^2$, in excellent agreement with the value obtained from the fit
to the $T_c$ vs. $d_S$ data.  Our independent estimate of $k_h$
varies over a broad range due to uncertainty in determining the
value of the diffusion constant (or mean free path) in
Ni.\cite{MoraruPRL2006} From the measured resistivity, we obtain
values of $l_F$ ranging between 7 and 70 nm, depending on what
value we take for the product $\rho_F l_F$ for
Ni.\cite{Niproperties}  Combining that with the values for the
exchange energy $E_{ex}=0.115$ eV and the Fermi velocity $v_F =
0.28 \times 10^6$ m/s,\cite{PetrovykhAPL1998} we obtain values for
$k_h$ ranging from $0.16-0.5$ nm$^{-1}$.  The value corresponding
to the shorter $l_F$ agrees with the value from the fit to the
data in Fig. 4.

The question remains open as to why Rusanov \textit{et al.}
\cite{RusanovPRB2006} observe inverse spin switch behavior, $T_c^P
> T_c^{AP}$, whereas we observe the standard behavior, $T_c^P
< T_c^{AP}$.  The most obvious difference between our samples and
theirs is that we use exchange bias to pin the magnetization
direction of one Py layer, whereas they rely on the different
coercivities of the two layers.  But the switching data in their
micron-scale samples show a clear plateau, which suggests that
they have achieved a good AP magnetization configuration.  A
second comment is that they observe a difference between $T_c^P$
and $T_c^{AP}$ even when the Nb layer is very thick, 60 nm,
whereas sensitivity to the ferromagnet orientation is limited to
our samples with $d_s<28$ nm. Variations in resistance or $T_c$
have also been observed in F/S bilayers due to domain formation
during magnetization
switching.\cite{RusanovPRL2004,KinseyIEEE2001} But Rusanov
\textit{et al.} state that the features indicating the inverse
spin switch effect in their trilayers were not observed in
bilayers. This fact, combined with their data in micron-scale
samples that appear to be single-domain, argue against any role of
domains in producing the inverse effect.

In summary, we observe similar spin-switch behavior in Py/Nb/Py
and Ni/Nb/Ni trilayers -- both S/F systems with strong
ferromagnets.  The results from both systems are fit well with the
dirty-limit Usadel theory of Fominov \textit{et al.}
\cite{FominovJETP2003} This success is somewhat unexpected given
that this oversimplified theory assumes identical electronic
properties (density of states, Fermi velocity, and mean free path)
for the majority and minority spin bands of the ferromagnetic
material, whereas Py is known to have a strong spin-scattering
asymmetry.  The success of a dirty-limit theory in Ni is also
surprising, and may be due partly to strong diffusive scattering
of electrons from the S/F interfaces.

We are grateful to J. Bass, R. Loloee and A. Rusanov for fruitful
discussions. This work was supported by NSF grants DMR 9809688,
0405238 and 0501013 and by the Keck Microfabrication Facility.


\begin{thebibliography} {99}
\bibitem{ThicknessDependence} In the thickness range of our data, the resistivity of bare Nb films
varies as $\rho_S$[n$\Omega$ m$] = 99 + 3415/d_s$[nm], while that
of the bare critical temperature is $T_{c\,0}$[K]$=9.1 -
43/d_s$[nm].
\end{thebibliography}
\end{document}